# On Deep Learning-based Massive MIMO Indoor User Localization

Maximilian Arnold, Sebastian Dörner, Sebastian Cammerer, Stephan ten Brink
Institute of Telecommunications, Pfaffenwaldring 47, University of Stuttgart, 70659 Stuttgart, Germany
{arnold,doerner,cammerer,tenbrink}@inue.uni-stuttgart.de

*Abstract*—We examine the usability of deep neural networks for multiple-input multiple-output (MIMO) user positioning solely based on the orthogonal frequency division multiplex (OFDM) complex channel coefficients. In contrast to other indoor positioning systems (IPSs), the proposed method does not require any additional piloting overhead or any other changes in the communications system itself as it is deployed on top of an existing OFDM MIMO system. Supported by actual measurements, we are mainly interested in the more challenging non-line of sight (NLoS) scenario. However, gradient descent optimization is known to require a large amount of data-points for training, i.e., the required database would be too large when compared to conventional methods. Thus, we propose a two-step training procedure, with training on simulated line of sight (LoS) data in the first step, and finetuning on measured NLoS positions in the second step. This turns out to reduce the required measured training positions and thus, reduces the effort for data acquisition.

## I. INTRODUCTION

Due to the huge success of mobile communication devices in almost any area of modern life, indoor positioning systems (IPSs) receive large attraction in both industry and academia. It can be seen as key enabler of a wide range of applications such as indoor navigation, smart factories, or could even provide a basic security functionality in distributed Internet of Things (IoT) sensor networks. Additionally, IPSs provide several promising benefits for existing technologies, e.g., for improved beamforming algorithms or motion prediction for channel estimation. On the other hand, multiple-input multiple-output (MIMO)-orthogonal frequency division multiplex (OFDM) systems are widely available and are workhorses of many state-of-the-art communication standards. Thus, it appears attractive to focus on such OFDM multi-antenna systems in the following.

While the line of sight (LoS) scenario is well-understood and multiple technologies are reported in the literature (e.g., angle- and time-of-arrival based predictions and triangulation methods) suitable solutions for the non-line of sight (NLoS) scenario are still open for research. Therefore, this work tackles the more challenging NLoS case which covers a huge variety of different scenarios. Although there exists an underlying channel transfer function which describes the behavior of the channel for any given position, this function is typically not known or can only be approximated as it is infeasible to fully capture the geometries of the room and its surrounding area. Thus, different approaches have been proposed (comp. [1], [2], [3], [4], [5], [6]) and investigated in the past, each optimized for different applications and system models (e.g., see [2] and [5] and references therein). Moreover, geomagnetic sensors were tested in [3] as an indoor positioning system in combination with deep learning. Overall, these approaches can be split into two categories, where obviously mixtures between both categories exist:

1) Model-based: define *how* the channel is expected to behave and estimate position accordingly (e.g., ray-tracing of a room).
2) Data-driven: collect a *database* with appropriate features (often called fingerprints) and corresponding positions (e.g., CSI [4], received signal strength indicator (RSSI) [7], [8] and recently time-reversal IPS (TRIPS) [9]) and, somehow, interpolate in-between.

In this work, we follow the data-driven approach, i.e., we do not assume any specific underlying channel model (except a basic pathloss-model in Sec. III). In comparison to most approaches in the literature (e.g., see [5]), we formulate the problem as a regression task rather than a classification problem, i.e., we directly predict 3D positions. Thus, the neural network implicitly has to learn a channel model or more precisely, it has to *approximate* the dependency between the channel coefficients and the corresponding positions. The authors of [4] follow a similar approach (however, for a convolutional neural network (CNN) structure) and report fractional-wavelength accuracy for a simulated channel model and thus, without considering the amount of training data.

In many applications, machine learning and, in particular deep learning, is known to be very effective whenever the underlying model is hard to aproxmiate or unknown (see [10]). Therefore, deep learning methods appear to be a good match for IPSs where the fundamental problem boils down to the question *How to approximate the channel behavior based only on a limited number of observations?* Unfortunately, training such a deep neural network typically requires a huge amount of training samples, hindering practical usability. As a result, besides the achievable accuracy, we believe the more fundamental question is how much information (datapoints) about a new environment does the system need in order to provides sufficient accuracy.

In the following, we propose a two-step training strategy, where the neural network (NN) is pre-trained on a simulated LoS channel in the first step, and later, in a second step finetuned, with only a small number of measured training samples. We then consider the performance of the NN for different reference case scenarios to answer basic questions on generalization and training strategies. We believe that the major contribution of this paper is to show that a NN, pre-trained on a simulated LoS channel, can learn the NLoS scenario faster (i.e., with less training data) than a randomly initialized NN. The intuition behind this approach is that both tasks are closely related. Thus, the weights from the previous task are a good starting point for learning the second task, leading to a reduction in overall training complexity. In machine learning terminology this means that we found a better way of initializing the weights of the NN, rather than a random initialization.

For the presented LoS and NLoS results all data was measured with a *spider-antenna* setup [11] which inherently provides ground truth (i.e., 3D position labels) at *sub-centimeter* precision.

## II. BACKGROUND

### A. Deep learning for user positioning

We refer the interested reader to [10] for a comprehensive introduction to machine learning. However for the sake of completeness and to clarify the notation we provide a short introduction.

In its basic form, a feed-forward NN is a directed computation graph consisting of multiple neurons with connections only to neurons of the subsequent layer. Each neuron sums up all weighted inputs and optionally applies a non-linear activation function, e.g., the rectified linear unit (ReLU) function

$$g_{\text{ReLU}}(x) = \max\{0, x\},$$

before forwarding the output to all connected neurons.

Let layer $i$ have $n_i$ inputs and $m_i$ outputs, then it performs the mapping $\mathbf{f}^{(i)} : \mathbb{R}^{n_i} \to \mathbb{R}^{m_i}$ defined by the weights and biases as parameters $\boldsymbol{\theta}_i$. Consecutively applying this mapping from input $\boldsymbol{v}$ of the first layer to the output $\boldsymbol{w}$ of the last layer, leads to the function

$$\boldsymbol{w} = \mathbf{f}\left(\boldsymbol{v}; \boldsymbol{\theta}\right) = \mathbf{f}^{(L-1)}\left(\mathbf{f}^{(L-2)}\left(\ldots\left(\mathbf{f}^{(0)}\left(\boldsymbol{v}\right)\right)\right)\right) \quad (1)$$

where $\boldsymbol{\theta}$ is the set of parameters and $L$ defines the *depth* of the net, i.e., the total number of layers.

*Training* of the NN describes the task of finding suitable weights $\boldsymbol{\theta}$ for a given dataset and its corresponding labels (desired output of the NN) such that a given loss function is minimized. This can be efficiently done with the stochastic gradient descent algorithm [10] as implemented in many state-of-the-art software libraries. We use the Tensorflow library [12]. In principle, the universal approximation theorem found in [13] states that such a multi-layer NN can approximate any continuous function on a bounded region arbitrarily closely for $L \geq 2$, given non-linear activation functions and a sufficiently large amount of neurons.

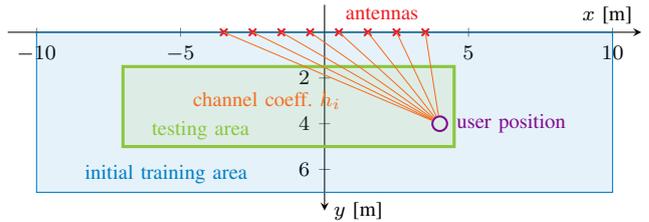

Fig. 1: System model under investigation.

### B. System Model

Fig. 1 shows the basic system model, where a linear antenna array (i.e 8 antennas on a line) is used. As indicated by the model, each antenna sees a slightly different single-tap complex channel coefficient $h_i$ for any user position and the NN needs to learn to estimate the user's position based on these inter-antenna channel differences. During the first training step the NN is initially trained on random user positions within the blue area in front of the antenna array. Since the 3D-spatial channel coefficients for any user position can be calculated by a LoS channel model, this training phase can be done with an unlimited amount of training data and with an arbitrarily large initial training area. The only limitation is that the initial training area should include that area which the NN is later tested and finetuned on. The initial training step is completed once the NN reaches a sufficient accuracy. An evaluation on the pre-trained NNs performance for different numbers of antennas and additive noise samples is shown in Section III.

The second training step uses measured datapoints taken from the green area (Fig. 1) to finetune the pre-trained NN for the actual testing area. In Section IV, we evaluate the required amount of finetuning datapoints and their respective positions. Note that this system model is only described for a 2-dimensional area, but can be extended to a 3D system straightforwardly. We use this system setup since all of our measurements are done in a 2D area with a 16-antenna linear array. Due to the antenna positions' symmetry, such a linear antenna array cannot distinguish between frontside and backside of the array. Therefore, a system that covers all positions of a 2D-plane also needs a 2-dimensional antenna array; accordingly for a 3D-area a 3-dimensional antenna array would be required.

*1) Simulated Channels:* The channel model can generate a complex channel coefficient $h_i$ for each spatial user position and antenna in a 3D area, as the (baseband) phase and amplitude of an LoS-channel can be computed using

$$h_{i,\text{LoS}} = \left(\frac{\lambda}{4\pi d}\right) e^{j2\pi \frac{d}{\lambda}}, \quad (2)$$

where $d$ is the distance between the transmitter antenna (TX) and the receiver antenna (RX), and $\lambda$ is the free-space wavelength. In this case a carrier frequency of $f_c = 2.35$ GHz is used. With this LoS channel model an infinite number of training points can be generated to pre-train the NN for the LoS scenario.

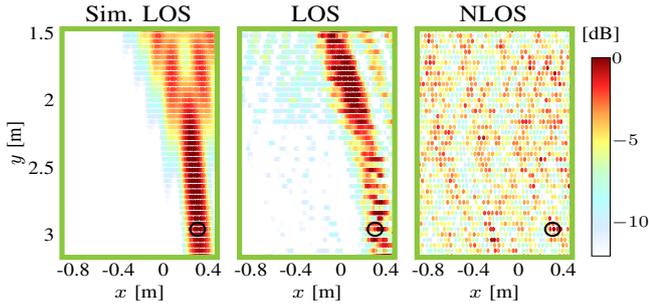

Fig. 2: Spatial energy map over testing area, sim. LoS vs meas. LoS vs meas. NLoS for MR precoding with a target user (black circle) at $x = 0.31$m, $y = 2.96$m, 16-antenna linear array.

*2) Measured Channels:* The spatial consistent channel measurements conducted in [11] will be briefly explained. A *spider antenna* is used to move a software-defined radio as a transmitter in the $x,y$-plane to measure a spatially consistent channel in two spatial dimensions, and to study linear transmit precoding. In Fig. 2, the spatial energy over the area is shown for the simulated LoS and the measured LoS and NLoS case for a spatial precoding: 16 antennas in a line with a distance of $\lambda/2$ are precoded with respect to the user position at $x=0.31$m and $y=2.96$m using a maximum ratio (MR) precoder. The single purpose of Fig. 2 is to show that there are only slight differences between the simulated and the measured LoS case, whereas there are huge differences to the NLoS case.

Remark: For this work, we only consider a single sub-carrier of the OFDM system per antenna. We have additionally trained NNs for all sub-carriers as input, but did not observe a significant gain besides the expected gain due to an improved SNR through averaging of the sub-carriers. However, the training complexity (and the required net dimension) increases drastically. The intuition behind is that the indoor channel (room size) does not show a significant frequency selective behavior for the measurement bandwidth of 40 MHz in our scenario (also not for the indoor NLoS case). Thus, the NN does not benefit from the increased amount of observations as there is only little additional information contained in sub-carrier interrelation and we conclude that one sub-carrier per antenna is sufficient.

## C. Performance metric

Due to the antennas' beam pattern, a user position with a large distance to the base station is more difficult to estimate than a user close to the base station. Therefore, we propose a performance metric to compare different measurement setups, algorithms, reference scenarios etc. which is independent of the user position. This metric is defined over the normalized mean squared error (NMSE) as a "relative" accuracy metric

$$\text{NMSE} = \mathbb{E}\left[\frac{\|\mathbf{p} - \hat{\mathbf{p}}\|^2}{\|\mathbf{p}\|^2}\right], \quad (3)$$

where the mean squared error (MSE) of the estimated position $\hat{\mathbf{p}}$ to the actual position $\mathbf{p}$ is normalized with respect to the distance $d = \|\mathbf{p}\|^2$ (assuming the antenna array to be at the origin and $\mathbf{p}$ is the user position in $x$-$y$-$z$ coordinates). For example, this metric returns a NMSE of 1% for a user position with a distance of 1m and an error distance of 1 cm. With increasing distance to the antenna array the difficulty of estimating the channel increases, which is compensated by this normalization.

## III. INITIAL TRAINING ON SIMULATED LOS CHANNELS

TABLE I: Layout of the neural net

| Layers: | Parameters | Output dimensions |
|---|---|---|
| Input | 0 | 16 (antennas) x 2 (Re/Im) |
| Dense (relu) | 33,792 | 1024 |
| Dense (relu) | 1,049,600 | 1024 |
| Dense (relu) | 1,049,600 | 1024 |
| Dense (linear) | 3075 | 3 |

In this section, we show a basic proof-of-concept, where the NN is able to approximate the user position, by evaluating the performance after the initial training step. Thus, the NN is tested on noisy, simulated LoS channel coefficients of random positions out of the blue area shown in Fig. 1. All results presented in the following have been achieved with a simple, non-hyperparameter-tuned NN, which has a total of 2,136,067 trainable weights and is defined in detail in Table I. Although the structure of the NN is quite simple as it only consists of dense layers, it contains a large amount of trainable parameters.[1] Thus, its capacity already exceeds our entire measured training data set (only $16 \times 2$ values per sample of about 60,000 measured positions), which is why we focused on preventing overfitting effects during finetuning. Note that the capacity of the NN is no problem during the pre-training stage, because we can train the NN with an infinite amount of different samples and thereby outnumber the NN's memory capabilities. This can be considered as another benefit of a pre-training stage, since it can obviate overfitting during the finetuning stage.

We introduce a signal-to-noise-ratio (SNR) definition

$$\text{SNR} = \frac{\sum_{n=1}^{N_{\text{ant}}} |h_{\text{LoS}}(n)|^2}{\sigma^2} \quad (4)$$

which is independent on the distance for all possible user positions. The notation, $\sigma^2$ is the power of the complex additive white Gaussian noise (AWGN), and $N_{\text{ant}}$ is the number of antennas at the base station, i.e., the linear array. Fig. 3 shows that a system with only 4 antennas can estimate the user position with an accuracy of 1% NMSE at an SNR of 30dB. Moreover, if the number of antennas is doubled, the expected gain of 3dB due to gains in noise averaging shows up. Note that, not the channel model function is learned by the NN but the *triangulation* task as a whole, which may even be more challenging. These results show that the NN is able to estimate the user position within the training area sufficiently well. But we also noticed that, if we test the NN on user positions outside of the initial training area, the estimations become rather bad. Apparently, the NN seems to learn an interpolation between fingerprints rather than it finds a global solution of the *triangulation* task itself. While this is not a problem for the simulated LoS channel scenario, since we can always generate the coefficients and train on all positions, it is a problem for real-world applications, as it implies that the NN needs to be trained on the whole area it is later used for and generalization to "external areas" is limited. This raises two key questions: *How many labeled samples are needed, and where do they have to be located?* For this, we define three reference scenarios for the second training step of finetuning on actual measured *data*, described in the following section.

---

[1]Remark: for the LoS scenario less weights are sufficient, however, the idea of *pre-training* requires the same dimensions for the LoS and NLoS scenario. Without considering complexity such a large structure turned out to reach higher accuracy for the NLoS case than smaller NNs. However, we cannot claim this observation to hold in general. Using CNNs may reduce the number of trainable parameters if required (cf. [4]), but also requires further hyper-parameter optimization.

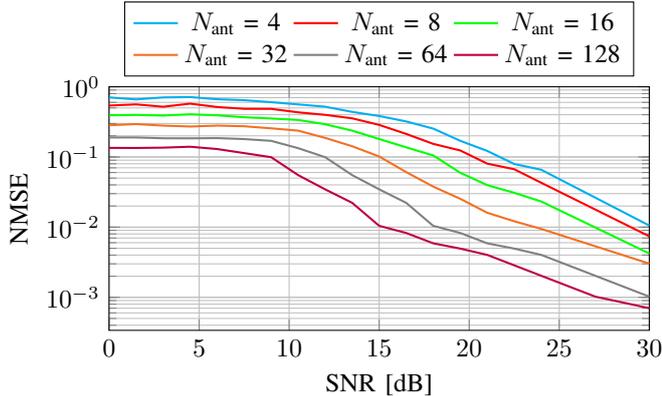

Fig. 3: Sweep over SNR and varying number of antennas for the simulated LoS scenario.

## IV. FINETUNING ON MEASURED LOS/NLOS DATA

Next, we consider the green area outlined in Fig. 1 and examine the required amount of datapoints and their spatial distribution for achieving a specific accuracy.

### A. Datapoint distribution

Fig. 4 schematically shows the three reference cases:
1) For the "random"-scenario, random datapoints taken from the whole testing area are used for finetuning training and, also, validation data to verify that the NNs accuracy is picked randomly from the whole testing area.
2) The "left-right" scenario is based on the idea of having trained a room (right side), but estimating on another room (left side), and splits the testing area into two areas where random datapoints from the left area are used for finetuning training and datapoints from the right area are used for validation.
3) In the "border"-scenario, which is motivated by the idea of only measuring the edges of a room, the finetuning datapoints are taken from a border area (here 30% of the area), and the validation datapoints are from the remaining (and unknown to the NN) center area.

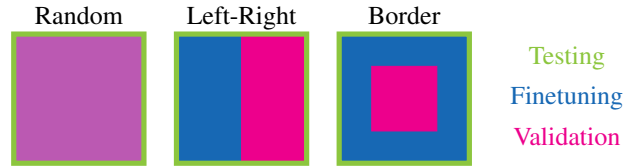

Fig. 4: Reference scenarios; different colors mark the different datasets.

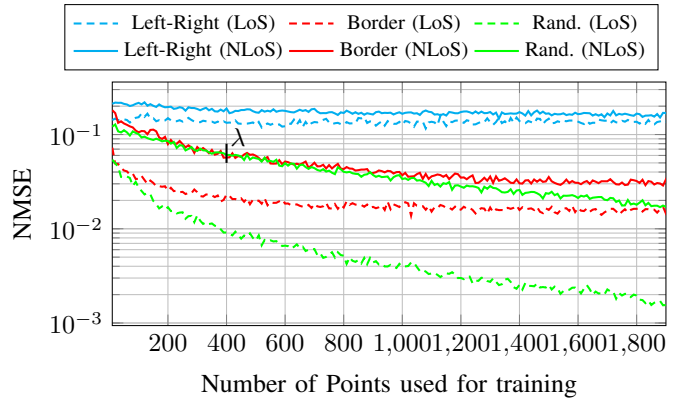

Fig. 5: NNs NMSE performance after finetuning on measured LoS (solid) and NLoS (dashed) area for different reference scenarios over the amount of used finetuning samples. 100 epochs are used for training.

Next, we use the NN, that has been pre-trained on the simulated LoS channel, (see previous section) and finetune it with actually measured LoS data according to the three different scenarios of Fig. 4. .

Fig. 5 compares the NMSE performance of the NN for the different scenarios in the LoS and NLoS case. Note that the random case is always used as the (best-case) reference. Apparently, for the "left-right" case, the NN does not converge at all, which matches with our observations for the simulated LoS channel. For the "border" case, the NN converges to a reasonably good estimation accuracy. It appears that the pre-trained NN only needs to see a few samples (about 400) from the border area during finetuning training to converge and estimate the positions reasonably well.

As expected, generally, the NLoS scenarios shows higher prediction errors caused by the more complex channel behavior. Similar to the LoS measurements, the NN does not converge in the "left-right" case, where the NMSE performance remains at about 10.8%. Interestingly, the performance of the "border" case performs similar to the random case up to about 1000 available finetuning samples. Intuitively, learning the channel transfer function for both the whole area, as well as for only the border area, is equally difficult since the channel exhibits a sort of random pattern in the NLoS case, as can be inferred from Fig. 2. After about 400 samples used for finetuning training, an accuracy of $\lambda$ is reached, which means that if each spatial $\lambda/2$ a new channel is assumed, the room has been fully learned by the NN. This shows that the pre-trained NN system can easily be adjusted and trained with low overhead and, therefore, is even a viable solution in the NLoS case.

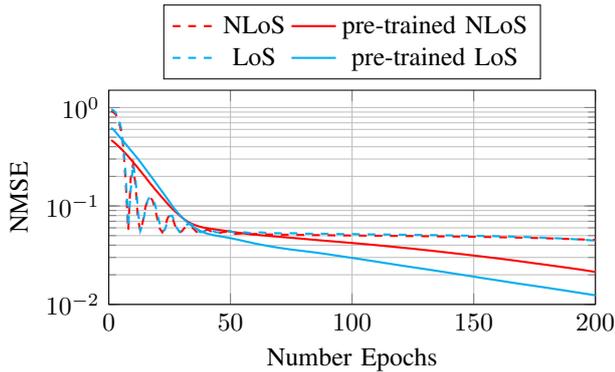

Fig. 6: Reachable NMSE over number of training epochs with 200 training samples between pre-trained and randomly initialized NN.

### B. Effect of pre-training

To demonstrate the benefit of pre-training on a simulated LoS channel, we compare the performance of the pre-trained NN with that of a randomly initialized NN in the context of the number of epochs needed to achieve a specific accuracy.

Fig. 6 shows the NMSE performance of a pre-trained NN and a randomly initialized NN trained on 200 measured samples for both the LoS and the NLoS case. Also the final NMSE reached by the pre-trained NN in both cases is lower than the final NMSE of the randomly initialized NN. To be fair, the randomly initialized NN can reach the NMSE performance of the pre-trained NN if it is trained with much more measured samples and/or epochs, but we consider the amount of available training data as a limiting factor of a real world application. This leads to the conclusion that pre-training of the NN on a simulated LoS channel is advantageous, as the pre-trained NN requires less measured and labeled training data and, also, converges faster for all scenarios tested.

## V. Outlook and Conclusion

In this work, we have shown that neural networks can be used for user localization in MIMO-OFDM systems. We have investigated the amount of required training positions and showed that pre-training of the NN with simulated LoS data significantly reduces the required amount of training samples. As a result, an accuracy of less than 1% for real data can be reached within our spatial test area. For future work a multi-room measurement setup will be build to verify the results in a wider and more general setting. The NN complexity (i.e., the number of trainable parameters) may be reduced by using additional CNN layers. Through standards such as WLAN 082.11n [14] MIMO-OFDM systems are widely available indoors and are an ideal candidate for systems where IPSs are useful. Thus, we believe the proposed system shines for its simplicity and comes with almost no additional cost.

## References

[1] A. Khalajmehrabadi, N. Gatsis, and D. Akopian, "Modern WLAN Fingerprinting Indoor Positioning Methods and Deployment Challenges," Oct. 2016.
[2] C. Chen, Y. Chen, Y. Han, H. Q. Lai, F. Zhang, and K. J. R. Liu, "Achieving centimeter-accuracy indoor localization on wifi platforms: A multi-antenna approach," *IEEE Internet of Things Journal*, vol. 4, no. 1, pp. 122–134, Feb 2017.
[3] N. Lee and D. Han, "Magnetic indoor positioning system using deep neural network," in *2017 International Conference on Indoor Positioning and Indoor Navigation (IPIN)*, Sept 2017, pp. 1–8.
[4] J. Vieira, E. Leitinger, M. Sarajlic, X. Li, and F. Tufvesson, "Deep Convolutional Neural Networks for Massive MIMO Fingerprint-Based Positioning," *ArXiv e-prints*, Aug. 2017.
[5] H. Chen, Y. Zhang, W. Li, X. Tao, and P. Zhang, "Confi: Convolutional neural networks based indoor wi-fi localization using channel state information," *IEEE Access*, vol. 5, pp. 18 066–18 074, 2017.
[6] X. Wang, L. Gao, and S. Mao, "Csi phase fingerprinting for indoor localization with a deep learning approach," *IEEE Internet of Things Journal*, vol. 3, no. 6, pp. 1113–1123, Dec 2016.
[7] P. Bahl and V. N. Padmanabhan, "Radar: an in-building rf-based user location and tracking system," in *Proceedings IEEE INFOCOM 2000. Conference on Computer Communications. Nineteenth Annual Joint Conference of the IEEE Computer and Communications Societies (Cat. No.00CH37064)*, vol. 2, 2000, pp. 775–784 vol.2.
[8] V. Savic and E. G. Larsson, "Fingerprinting-based positioning in distributed massive mimo systems," in *2015 IEEE 82nd Vehicular Technology Conference (VTC2015-Fall)*, Sept 2015, pp. 1–5.
[9] Z. H. Wu, Y. Han, Y. Chen, and K. J. R. Liu, "A time-reversal paradigm for indoor positioning system," *IEEE Transactions on Vehicular Technology*, vol. 64, no. 4, pp. 1331–1339, April 2015.
[10] I. Goodfellow, Y. Bengio, and A. Courville, *Deep Learning*. MIT Press, 2016.
[11] M. Arnold, M. Gauger, and S. ten Brink, "Evaluating massive MIMO precoding based on 3D-channel measurements with a spider antenna," in *2017 International Symposium on Wireless Communication Systems (ISWCS)*, Aug 2017, pp. 134–139.
[12] M. A. et al., "Tensorflow: Large-scale machine learning on heterogeneous distributed systems," *CoRR*, 2016. [Online]. Available: http://arxiv.org/abs/1603.04467
[13] K. Hornik, M. Stinchcombe, and H. White, "Multilayer feedforward networks are universal approximators," *Neural Networks*, vol. 2, no. 5, pp. 359–366, 1989.
[14] J. Lorincz and D. Begusic, "Physical layer analysis of emerging ieee 802.11n wlan standard," in *2006 8th International Conference Advanced Communication Technology*, vol. 1, Feb 2006, pp. 6 pp.–194.